\documentclass{article}
\usepackage{spconf,amsmath,graphicx,hyperref}
\usepackage{algorithm}
\usepackage{algpseudocode}
\usepackage{float}
\usepackage{amsthm}
\usepackage{booktabs}
\usepackage[table]{xcolor}

\title{Pianoroll-Event: A Novel Score Representation for Symbolic Music}
\name{Lekai Qian$^\ast$, Haoyu Gu$^\ast$, Dehan Li$^\ast$, Boyu Cao$^\dagger$, and Qi Liu$^\dagger$\thanks{$^\ast$Equal contribution. $^\dagger$Corresponding authors. This work was supported in part by the National Natural Science Foundation of China under Grant 62202174, in part by the GJYC program of Guangzhou under Grant 2024D01J0081, in part by the ZJ program of Guangdong under Grant 2023QN10X455, and in part by the Fundamental Research Funds for the Central Universities under Grant 2025ZYGXZR053.}}
\address{School of Future Technology, South China University of Technology}
\begin{document}
%
\maketitle
\begin{abstract}
Symbolic music representation is a fundamental challenge in computational musicology. While grid-based representations effectively preserve pitch-time spatial correspondence, their inherent data sparsity leads to low encoding efficiency. Discrete-event representations achieve compact encoding but fail to adequately capture structural invariance and spatial locality. To address these complementary limitations, we propose Pianoroll-Event, a novel encoding scheme that describes pianoroll representations through events, combining structural properties with encoding efficiency while maintaining temporal dependencies and local spatial patterns. Specifically, we design four complementary event types: Frame Events for temporal boundaries, Gap Events for sparse regions, Pattern Events for note patterns, and Musical Structure Events for musical metadata. Pianoroll-Event strikes an effective balance between sequence length and vocabulary size, improving encoding efficiency by 1.36$\times$ to 7.16$\times$ over representative discrete sequence methods. Experiments across multiple autoregressive architectures show models using our representation consistently outperform baselines in both quantitative and human evaluations.
\end{abstract}
\section{Introduction}

Symbolic music representation provides the foundation for computational musicology, yet the inherent complexity of musical structures poses significant challenges for establishing efficient encoding schemes. In contrast to natural language processing and computer vision with their standardized representations, the symbolic music community lacks consensus on optimal encoding strategies.

Current research in symbolic music representation predominantly follows two paradigms: continuous-time representation and discrete-event representation. Continuous-time approaches (e.g., pianoroll~\cite{boulanger2012modeling}) preserve the spatial relationship between pitch and temporal positions by modeling symbolic music as 2D matrices. However, the inherent sparsity of musical events typically results in matrices with substantial empty values, which may compromise the efficiency of information encoding. Discrete-event representations encode music as token sequences (e.g., ABC notation~\cite{walshaw1996abc}, MIDI Events~\cite{mma1996midi}), facilitating efficient modeling. While generally offering improved information efficiency, such approaches may not adequately capture inherent musical invariances, as critical properties like relative pitch intervals are often encoded as absolute values.

To address the limitations of existing representation paradigms, we propose Pianoroll-Event, a novel encoding scheme that combines the spatial structure preservation of pianoroll with the efficiency of discrete-event sequential representations. Our method first discretizes pianoroll into temporal frames to preserve time-dependent relationships, then transforms these frames into structured event sequences through four complementary event types: Frame Events that mark frame boundaries while compressing consecutive empty pitch regions, Pattern Events that encode local note activation patterns, Gap Events that efficiently represent sparse pitch intervals within frames, and Musical Structure Events that capture essential musical context including bar positions and time signature changes. Through this design, Pianoroll-Event achieves effective music information compression while ensuring each tokenized element maintains clear semantic meaning. In summary, the contributions of this paper are as follows:
\begin{itemize}
\item We propose Pianoroll-Event, a novel symbolic music encoding that bridges spatial and sequential representations by designing four complementary event types to transform pianoroll into efficient token sequences.
\item We analyze encoding efficiency in terms of sequence length and vocabulary size, demonstrating significant computational advantages with compression ratios ranging from 1.36$\times$ to 7.16$\times$ over existing approaches.
\item We conduct comprehensive experiments showing that Pianoroll-Event achieves state-of-the-art performance with improvements ranging from 3.43\% to 47.00\% in objective metrics and 30.61\% to 66.56\% in subjective evaluations across mainstream sequence modeling architectures.
\end{itemize}

\section{Related Work}
\subsection{Symbolic Music Representation Methods}
Symbolic music representation serves as the foundation of computational musicology, requiring careful balance among information density, computational efficiency, and expressive capability~\cite{briot2020deep,ji2023survey,zhang2023symbolic}. \textbf{Pianoroll representation}~\cite{boulanger2012modeling} encodes music as a two-dimensional matrix where the vertical axis represents pitch and the horizontal axis represents time. However, as analyzed by Walder~\cite{walder2016modelling}, piano rolls suffer from inherent sparsity that leads to computational redundancy, and lack explicit note boundary information. \textbf{Event-based representations} encode music as discrete token sequences. ABC Notation~\cite{walshaw1996abc} provides human-readable format but is primarily designed for monophonic melodies. MIDI event sequences~\cite{mma1996midi} precisely record note attributes, though early methods~\cite{oore2018thistimewithfeeling} preserve complete information at the cost of excessive sequence length. REMI~\cite{huang2020popmusictransformer} reduces sequence length through beat-relative position markers. Compound Word~\cite{hsiao2021compoundword} groups related attributes into compound tokens, while OctupleMIDI~\cite{zeng2021musicbert} encodes attributes as octuple tokens. Byte Pair Encoding~\cite{fradet2023bpe} learns frequent patterns for compression.

\subsection{Sequence Modeling and Autoregressive Generation}
Sequence modeling has become dominant in symbolic music generation. LSTM~\cite{hochreiter1997lstm} mitigates long-term dependencies through gating mechanisms. The Transformer~\cite{vaswani2017attention} achieves global dependency modeling via self-attention, with Music Transformer~\cite{huang2018music} introducing relative position encoding. Pop Music Transformer~\cite{huang2020popmusictransformer} and MuseNet~\cite{payne2019musenet} further advance the field through extended context and large-scale pretraining. These advances underscore the interplay between representation design and generation quality.
\section{Method}

\begin{figure*} [h]
    \centering
    \includegraphics[width=0.8\linewidth]{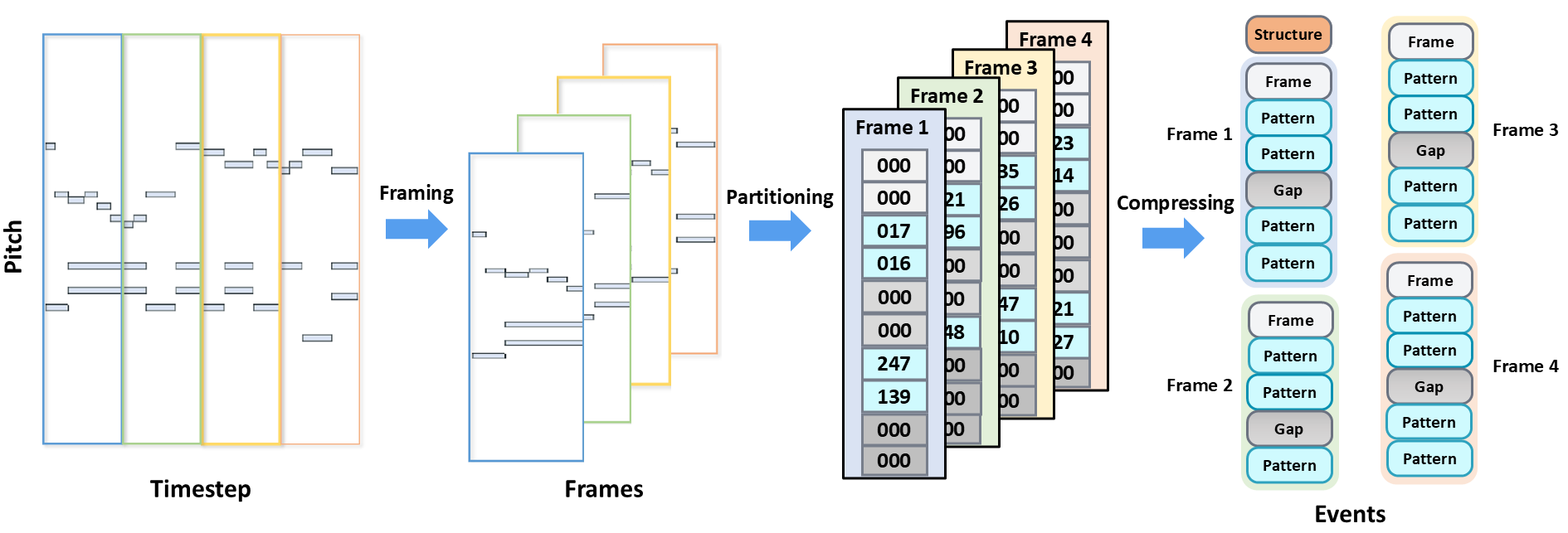}
    \caption{The process of converting pianoroll representation into pianoroll-events. Through frame segmentation, partitioning, and compression operations, the pianoroll is transformed into a sequence of pianoroll-events containing diverse event types.}
    \label{fig:pipeline}
\end{figure*}

This section describes the Pianoroll-Event conversion process, as illustrated in Fig.~\ref{fig:pipeline}. First, we partition the pianoroll into fixed-length frames. Then, we convert each frame into a sequence of four events that efficiently encode the sparse pitch information. Finally, we tokenize these events to produce sequences compatible with standard sequence modeling architectures.

\subsection{Temporal Framing}

We apply temporal framing to the pianoroll to generate fundamental units for subsequent transformation. Let $\mathbf{P} \in \{0,1\}^{H \times T}$ denote a pianoroll, where $H=88$ represents the standard piano pitch range and $T$ is the number of time steps. The framing process partitions $\mathbf{P}$ into an ordered sequence $\{\mathbf{F}_1, \mathbf{F}_2, \ldots, \mathbf{F}_N\}$, with $N = \lceil T/L \rceil$ and each frame $\mathbf{F}_i \in \{0,1\}^{H \times L}$. Specifically, each frame is obtained as:

\begin{equation}
\mathbf{F}_i = \mathbf{P}[:, (i-1)L : \min(iL, T)]
\end{equation}

where the notation $\mathbf{P}[:, a:b]$ denotes the slice of $\mathbf{P}$ from time step $a$ to $b-1$. This procedure preserves local musical structures like chordal verticality and melodic continuity, and maintains strict temporal dependencies across frames, enabling downstream models to capture music's temporal logic.

\subsection{Pianoroll-Event Generation}

A fundamental characteristic of pianoroll representations is their inherent sparsity—most entries are zeros, with active notes typically concentrated within limited pitch ranges. This sparsity motivates our event-based encoding strategy, which transforms dense frame representations into compact event sequences. We uniformly partition each frame $\mathbf{F}_i$ along the pitch dimension into fixed-size blocks, then encode these blocks as typed events based on their content and position. 

Specifically, we partition each frame $\mathbf{F}_i \in \{0,1\}^{H \times L}$ into $K = \lceil H/h \rceil$ Event Blocks along the pitch dimension via direct slicing: $\mathbf{B}_{i,j} = \mathbf{F}_i[(j-1)h : jh, :]$, where each block $\mathbf{B}_{i,j} \in \{0,1\}^{h \times L}$ is a contiguous sub-matrix capturing $h$ consecutive pitches. The compression stage then examines these blocks sequentially to generate typed events. The consecutive empty blocks at the beginning of a frame are merged into a single Frame Event, which marks the frame boundary and encodes the start position. Conversely, the consecutive empty blocks at the end are discarded, as they can be reconstructed during decoding given the fixed frame length. For the intermediate blocks, each non-empty block $\mathbf{B}_{i,j}$ is mapped to a unique Pattern Event: $\text{Pattern}(\mathbf{B}_{i,j})$ that preserves its note configuration, while a sequence of $r$ consecutive empty blocks is compressed into a single Gap Event: $\text{Gap}(r)$. Additionally, Musical Structure Events are inserted at measure boundaries to encode time signatures, bar lines, and other symbolic information.

These events are then mapped to discrete tokens for sequence modeling. Each event type corresponds to a distinct token vocabulary: Frame tokens encode starting positions, Pattern tokens represent unique note configurations, Gap tokens specify run lengths, and Musical Structure tokens encode symbolic elements. Given a pianoroll $\mathbf{P} \in \{0,1\}^{H \times T}$ partitioned into frames $\{\mathbf{F}_1, \mathbf{F}_2, \ldots, \mathbf{F}_N\}$, the final event sequence is obtained as:

\begin{equation}
\mathbf{S} = \bigoplus_{i=1}^{N} \left[ \text{Encode}(\mathbf{F}_i) \oplus \text{MSE}_i \right]
\end{equation}
where $\text{Encode}(\cdot)$ denotes the frame encoding process described in Algorithm~\ref{alg:pianoroll_event}, $\text{MSE}_i$ represents optional Musical Structure Events at measure boundaries, and $\oplus$ denotes concatenation. This semantic mapping provides downstream models with meaningful token-level interpretations while achieving substantial compression. The resulting sequences preserve temporal structure and pitch patterns, maintaining compatibility with standard sequence modeling architectures.


\begin{algorithm}[ht]
\caption{Pianoroll-Event Encoding Process}\label{alg:pianoroll_event}
\begin{algorithmic}[1]
\Require Pianoroll frame $\mathbf{F} \in \{0,1\}^{H \times L}$
\Ensure Event sequence $\mathbf{S}$
\State Partition $\mathbf{F}$ into blocks $\{\mathbf{B}_1, \mathbf{B}_2, \ldots, \mathbf{B}_K\}$ of size $h \times L$
\State Merge leading empty blocks into $\text{Frame\_Event}$
\State $\mathbf{S} \gets [\text{Frame\_Event}]$; $r \gets 0$
\For{each block $\mathbf{B}_i$ in the middle region}
    \If{$\mathbf{B}_i$ is empty}
        \State $r \gets r + 1$ 
    \Else
        \State \textbf{if} $r > 0$: $\mathbf{S}.\text{append}(\text{Gap\_Event}(r))$; $r \gets 0$
        \State $\mathbf{S}.\text{append}(\text{Pattern\_Event}(\mathbf{B}_i))$
    \EndIf
\EndFor
\State \textbf{if} $r > 0$: $\mathbf{S}.\text{append}(\text{Gap\_Event}(r))$
\State Drop trailing consecutive empty blocks
\State \Return $\mathbf{S}$
\end{algorithmic}
\end{algorithm}

\section{Experiments and Analysis}

\subsection{Dataset}

We use the MuseScore dataset containing 140,000 two-track piano scores with durations ranging from 1 to 5 minutes. We convert these scores into multi-hot array piano roll representations with a temporal resolution 1/16 of a beat. Each time step preserves 88 pitch values corresponding to the standard piano keyboard. We then encode the piano rolls into different token representations for training sequence models.

\subsection{Encoding Efficiency Analysis}

We evaluate encoding efficiency using the Budget-Aware Difficulty Index (BDI), defined as:
\begin{equation}
\text{BDI} = \ell^2 \times \sqrt{V}
\end{equation}
where $\ell$ is the average sequence length and $V$ is the vocabulary size. This metric captures both the quadratic computational complexity of self-attention mechanisms and the capacity dilution effect of large vocabularies on model parameters.

Table~\ref{tab:encoding_efficiency} compares five encoding methods. Our approach achieves the lowest BDI ($1.048\times10^7$), showing an optimal balance between sequence compression and vocabulary size. For fair comparison, we exclude velocity tokens from all sequence-based methods. Yet our approach still maintains its efficiency advantage over both traditional long-sequence methods and vocabulary-heavy BPE approaches.

\begin{table}[h]
\centering
\caption{Encoding efficiency comparison}
\label{tab:encoding_efficiency}
\begin{tabular}{lcccc}
\toprule
\textbf{Method} & \textbf{$\ell$} & \textbf{$V$} & \textbf{BDI} $\downarrow$ & \textbf{vs. Ours} $\downarrow$ \\
\midrule
\textbf{Ours} & 749.8 & 347 & \textbf{1.048} & \textbf{1.00$\times$} \\
REMI & 1339.7 & 330 & 3.261 & 3.11$\times$ \\
MIDILike & 1398.9 & 448 & 4.143 & 3.96$\times$ \\
REMI-BPE & 317.8 & 20,000 & 1.429 & 1.36$\times$ \\
ABC Notation & 2575.0 & 128 & 7.504 & 7.16$\times$ \\
\bottomrule
\end{tabular}
\end{table}

\subsection{Generation Quality Evaluation}

We train all models for 20 epochs on the MuseScore dataset using an NVIDIA RTX 4090, with learning rate 1e-4 and batch size 256. The model configurations are: GPT-2-Small (4 layers, 512 hidden size), GPT-2-Large (8 layers, 768 hidden size), Llama (6 layers, 768 hidden size), and LSTM (4 layers, 512 hidden size). For evaluation, we generate 50 samples per method targeting 40-90 second durations for statistical analysis. Subjective assessment involves 10 samples per model evaluated by 30 musically trained listeners in a double-blind setup, with scores averaged to obtain the final \textbf{MOS}.

We employ objective metrics from MusPy~\cite{dong2020muspy}: \textbf{Polyphony Rate (PR)} for harmonic richness, \textbf{Groove Consistency (GC)} for rhythmic stability, and \textbf{Scale Consistency (SC)} for tonal coherence. The \textbf{JS Divergence Similarity} is computed as $\text{JS} = 100 \times \exp(-2 \times \overline{\text{JS}})$, where $\overline{\text{JS}}$ represents the average Jensen-Shannon divergence calculated using mean and variance of the distributions across all three metrics. Tables~\ref{tab:gpt2_small}, \ref{tab:gpt2_large}, \ref{tab:llama}, and \ref{tab:lstm} show the evaluation results with mean values.
\begin{table}[h]
\centering
\caption{GPT-2-Small Model Experimental Results}
\label{tab:gpt2_small}
\begin{tabular}{lccccc}
\toprule
\textbf{Method} & \textbf{PR} & \textbf{GC} & \textbf{SC} & \textbf{JS}$\uparrow$ & \textbf{MOS}$\uparrow$ \\
\midrule
REMI & 0.735 & 0.844 & 0.698 & 34.86 & 1.97 \\
REMI-BPE & 0.612 & 0.847 & 0.844 & 54.36 & 2.50 \\
MIDI-Event & 0.773 & 0.869 & 0.692 & 39.33 & 2.07 \\
CP & 0.634 & 0.773 & 0.807 & 51.08 & 1.83 \\
Octuple & 0.041 & 0.928 & 0.905 & 43.31 & 1.67 \\
ABC & 0.398 & 0.997 & 0.977 & 63.29 & 2.63 \\
\textbf{Ours} & 0.653 & 0.927 & 0.917 & \textbf{67.95} & \textbf{3.80} \\
\midrule
GT & 0.583 & 0.980 & 0.943 & - & 4.83 \\
\bottomrule
\end{tabular}
\end{table}

\begin{table}[h]
\centering
\caption{GPT-2-Large Model Experimental Results}
\label{tab:gpt2_large}
\begin{tabular}{lccccc}
\toprule
\textbf{Method} & \textbf{PR} & \textbf{GC} & \textbf{SC} & \textbf{JS}$\uparrow$ & \textbf{MOS}$\uparrow$ \\
\midrule
REMI & 0.751 & 0.992 & 0.710 & 35.85 & 1.07 \\
REMI-BPE & 0.286 & 0.815 & 0.878 & 55.27 & 2.93 \\
MIDI-Event & 0.748 & 0.855 & 0.709 & 40.53 & 2.03 \\
CP & 0.719 & 0.726 & 0.799 & 49.93 & 3.00 \\
Octuple & 0.078 & 0.916 & 0.909 & 50.61 & 2.33 \\
ABC & 0.261 & 0.997 & 0.966 & 65.18 & 2.00 \\
\textbf{Ours} & 0.742 & 0.936 & 0.962 & \textbf{68.86} & \textbf{4.27} \\
\midrule
GT & 0.583 & 0.980 & 0.943 & - & 4.83 \\
\bottomrule
\end{tabular}
\end{table}

\begin{table}[h]
\centering
\caption{Llama model experimental results}
\label{tab:llama}
\begin{tabular}{lccccc}
\toprule
\textbf{Method} & \textbf{PR} & \textbf{GC} & \textbf{SC} & \textbf{JS}$\uparrow$ & \textbf{MOS}$\uparrow$ \\
\midrule
REMI & 0.805 & 0.804 & 0.695 & 34.87 & 1.13 \\
REMI-BPE & 0.346 & 0.860 & 0.832 & 49.94 & 2.07 \\
MIDI-Event & 0.955 & 0.835 & 0.719 & 31.56 & 2.07 \\
CP & 0.715 & 0.749 & 0.782 & 45.46 & 1.20 \\
Octuple & 0.111 & 0.918 & 0.853 & 45.58 & 1.36 \\
ABC & 0.368 & 0.997 & 0.960 & 64.77 & 4.33 \\
\textbf{Ours} & 0.668 & 0.912 & 0.959 & \textbf{64.94} & \textbf{4.67} \\
\midrule
GT & 0.583 & 0.980 & 0.943 & - & 4.83 \\
\bottomrule
\end{tabular}
\end{table}

\begin{table}[h]
\centering
\caption{LSTM model experimental results}
\label{tab:lstm}
\begin{tabular}{lccccc}
\toprule
\textbf{Method} & \textbf{PR} & \textbf{GC} & \textbf{SC} & \textbf{JS}$\uparrow$ & \textbf{MOS}$\uparrow$ \\
\midrule
REMI & 0.424 & 0.991 & 0.709 & 34.48 & 1.27 \\
REMI-BPE & 0.366 & 0.811 & 0.857 & 39.90 & 3.00 \\
MIDI-Event & 0.770 & 0.821 & 0.878 & 56.25 & 1.73 \\
CP & 0.750 & 0.806 & 0.689 & 35.63 & 1.90 \\
Octuple & 0.167 & 0.912 & 0.948 & 56.61 & 2.43 \\
ABC & 0.281 & 0.975 & 0.957 & 61.97 & 2.33 \\
\textbf{Ours} & 0.601 & 0.937 & 0.832 & \textbf{62.53} & \textbf{3.53} \\
\midrule
GT & 0.583 & 0.980 & 0.943 & - & 4.83 \\
\bottomrule
\end{tabular}
\end{table}

Our method consistently achieves the highest JS scores across all architectures, with GC and SC metrics maintaining values above 0.93 and 0.92 respectively. This demonstrates superior alignment with ground truth distributions compared to all baseline methods.

\subsection{Ablation Study}

We conduct ablation experiments on GPT-2-Large to evaluate the contribution of each encoding component. Starting from pattern events only (P), we progressively add frame compression for leading zeros (PF+), remove trailing zeros (PF), and finally incorporate gap tokens for internal zeros (Proposed). As shown in Table~\ref{tab:ablation}, both objective metrics and subjective evaluations demonstrate consistent improvement across variants. The JS similarity increases from 50.16 (P) to 68.86 (Proposed), while MOS improves from 2.20 to 4.07, confirming that each component contributes meaningfully to the encoding's effectiveness.

\begin{table}[h]
\centering
\caption{Ablation study on GPT-2-Large model}
\label{tab:ablation}
\begin{tabular}{lccccc}
\toprule
\textbf{Method} & \textbf{PR} & \textbf{GC} & \textbf{SC} & \textbf{JS}$\uparrow$ & \textbf{MOS}$\uparrow$ \\
\midrule
P & 0.370 & 0.723 & 0.714 & 50.16 & 2.20 \\
PF+ & 0.683 & 0.905 & 0.962 & 60.92 & 3.20 \\
PF & 0.716 & 0.900 & 0.945 & 62.96 & 3.67 \\
Proposed & 0.742 & 0.936 & 0.962 & \textbf{68.86} & \textbf{4.07} \\
\midrule
GT & 0.583 & 0.980 & 0.943 & - & 4.8 \\
\bottomrule
\end{tabular}
\end{table}

\section{Conclusion}
We presented Pianoroll-Event, a novel symbolic music encoding that bridges grid-based and discrete-event representations through four complementary event types. Our approach preserves spatial-temporal relationships while achieving superior encoding efficiency compared to existing methods. Experiments demonstrate state-of-the-art generation quality across various model architectures in both objective and subjective evaluations. This work provides a promising unified representation framework for advancing computational musicology and music generation tasks.

\vfill\pagebreak

\bibliographystyle{IEEEbib}
\bibliography{strings,refs}

\end{document}